\documentclass[twocolumn]{aastex61}				% Linux

\newcommand\aastex{AAS\TeX}

\received{2018 May 8}
\revised{2018 July 9}
\accepted{2018 July 9}
\submitjournal{ApJ}

\shorttitle{\aastex\ The low detection rate of pair instability supernovae}
\shortauthors{Takahashi, K.}

\begin{document}

\title{The low detection rate of pair instability supernovae and the effect of the core carbon fraction}

\correspondingauthor{Koh Takahashi}
\email{ktakahashi@astro.uni-bonn.de}

\author{Koh Takahashi}
\affil{Argelander-Institut f\"{u}r Astronomie, Universit\"{a}t Bonn,
D-53121 Bonn, Germany}

\begin{abstract}
The pair instability supernova (PISN) is a common fate of very massive stars (VMSs).
Current theory predicts the initial and the CO core mass ranges for PISNe
of $\sim$140--260 $M_\odot$ and $\sim$65--120 $M_\odot$ respectively
for stars that are not much affected by the wind mass loss.
The corresponding relative event rate between PISNe and core collapse supernovae 
is estimated to be $\sim$1\% for the present-day initial mass function.
However, no confident PISN candidate has been detected so far,
despite more than 1,000 supernovae are discovered every recent years.
We investigate the evolution of VMSs with various core carbon-to-oxygen ratios for the first time,
by introducing a multiplication factor $f_{\rm cag} \in [ 0.1, 1.2 ]$
to the $^{12}$C($\alpha, \gamma$)$^{16}$O reaction rate.
We find that a less massive VMS with a high $X$(C)/$X$(O) 
develops shell convection during the core carbon-burning phase,
with which the star avoids the pair-creation instability.
The second result is the high explodability for a massive VMS, i.e.,
a star with high $X$(C)/$X$(O) explodes with a smaller explosion energy.
Consequently, the initial and the CO core mass ranges for PISNe are significantly increased.
Finally, a PISN with high $X$(C)/$X$(O) yields smaller amount of $^{56}$Ni.
Therefore, PISNe with high $X$(C)/$X$(O) are much rarer and fainter to be detected.
This result advances the first theory to decrease the PISN event rate 
by directly shifting the CO core mass range.

\end{abstract}

\section{Introduction}

The pair instability supernova (PISN) is known as a common fate of very massive
stars (VMSs)\footnote{Very massive stars with the initial mass of $>100$ $M_\odot$
are conventionally called as VMSs  \citep[e.g.,][]{Vink+15}.}
that develop massive CO cores during the evolution \citep{barkat+67, rakavy+67}.
In the massive CO core, the $e^- e^+$ pair creation reaction effectively takes place,
converting the thermal energy into the rest mass of $e^- e^+$ pair and softening the pressure.
As a consequence, the core becomes hydrodynamically unstable
and initiates accelerating core contraction, or, core collapse.
In the collapsing core, nuclear reactions of carbon, neon, and oxygen burnings take place.
If the nuclear reactions release large energy enough to explode the whole star,
then the explosion called PISN takes place in the end.
Indeed, hydrodynamical simulations in both
1D \citep[e.g.,][]{heger&woosley02, umeda&nomoto02, kozyreva+14a, takahashi+16} and
multi-D \citep{chatzopoulos+13b, chen+14a} have resulted in successful explosions,
confirming the robustness of the mechanism.
The confident understanding of the explosion mechanism provides
a strong motivation to search a PISN explosion in the real universe.

In spite of the robust theoretical prediction, existence of PISNe
has not been observationally confirmed so far.
The explosion of a PISN can be observed as a luminous supernova
because of the large explosion energy and the large $^{56}$Ni yield \citep[e.g.][]{kasen+11},
while lower mass PISNe are expected to be dim \citep{kasen+11, kozyreva+14a}.
Therefore a class of so-called super luminous supernovae (SLSNe),
which shows a luminosity 10 or more times larger than that of a standard supernova,
is a good candidate to be explained as a PISN event \citep{gal-yam+09, gal-yam12}.
However, no currently observed SLSNe match with theoretical predictions,
which produce much broader light curves and more red colors than observations
as a result of the intrinsically long diffusion timescale of the large ejecta masses
\citep{dessart+12, dessart+13, chatzopoulos+15, kozyreva+14b, kozyreva+16, kozyreva+17}.
Besides existence of PISNe in the early universe has not been confirmed yet.
Instead of the direct detection, PISN explosions in the early universe can be traced
by observing surface chemical abundances of metal-poor stars
\citep[abundance profiling;][]{nomoto+13,umeda&nomoto02}.
Although thousands of metal-poor stars have been observed until now,
none of them show agreements with PISN characteristic abundances of
low [Na/Mg]\footnote{A stellar abundance ratio is indicated by the solar scaled value of
$[X/Y] \equiv {\rm log} ( n_X/n_Y ) - {\rm log} ( n_{X,\odot}/n_{Y,\odot} )$,
where $n_X$ is the number density of the element $X$ and $n_{X,\odot}$ is the solar value.}
and high [Ca/Mg] \citep{takahashi+18a}.

Stellar evolution simulations of single nonrotating zerometallicity VMSs have estimated
the mass range of CO cores for PISNe to be $\sim$65--120 $M_\odot$ and 
the corresponding initial mass range of $\sim$140--260 $M_\odot$ 
\citep{heger&woosley02, takahashi+16}.
Assuming the Salpeter initial mass function (IMF) with a slope of $\alpha=2.35$,
a relative number fraction of PISNe to core collapse supernovae (CCSNe) 
of $\sim$1\% is estimated.
The relative event rate between PISNe and CCSNe 
can not be directly related to the detection rate
because these supernovae will have various luminosities and durations
and supernova surveys are magnitude and volume limited.
Nevertheless, provided that more than 1,000 of supernovae are discovered every year 
by the current supernova surveys (the “Latest Supernovae” 
website\footnote{http://www.rochesterastronomy.org/supernova.html}; 
\citealt{Gal-Yam+13}, and references therein), 
the $\sim$1\% relative event rate might be 
large enough for the PISN detection.
Furthermore, the number of detection will be significantly increased
after the operation start of upcoming surveys such as
the Zwicky Transient Facility (ZTF\footnote{http://www.ztf.caltech.edu/}, \citealt{Bellm18})
and the Large Synoptic Survey Telescope (LSST\footnote{https://www.lsst.org/}, \citealt{Ivezic+08}).

If the PISN confident detection will not be achieved even by the upcoming surveys,
it will imply that the actual event rate is well below $\sim$1\% obtained by the present theory.
One important uncertainty related to the event rate estimate
is involved in the upper limiting mass for the star formation.
If the limiting mass is below the lower end of the PISN mass range,
no PISN takes place in the universe.
However, the estimated initial mass of $\gtrsim$ 140 $M_\odot$ for a PISN progenitor
is not so much massive as regarded as unrealistic.
A VMS with the initial mass of $\lesssim$ 320 $M_\odot$
has been observed in a star cluster R136 in the Large Magellanic Cloud \citep{crowther+10, crowther+16},
and the upper limiting mass for the star formation in the cluster
has been estimated to be $\gtrsim$200 $M_\odot$ \citep{Schneider+14},
indicating that the formation of a VMS is possible for environments with finite metallicities.
Apart from that, the star formation in the early universe 
has been investigated by cosmological ab-initio simulations.
Because of the absence of efficient coolants in primordial gas clouds,
zerometallicity stars are considered to be born with very massive initial masses
of $\sim$100 $M_\odot$ \citep[][and references therein]{hirano+14, hirano+15, susa+14}.

The other big uncertainties are in the estimate of the initial mass range for PISNe. 
If the present theory has estimated 
the lower minimum initial mass for a PISN than the actual value,
then it overestimates the event rate
because a less massive star is more frequently formed
under the present initial mass function.
One of the relevant physics is the strong wind mass loss, 
and especially its metallicity dependence is of importance.
Due to the strong wind during both the main sequence and the Walf-Rayet phases,
a solar metallicity VMS of $\lesssim$500 $M_\odot$ will not become a PISN \citep{yusof+13, yoshida+14}.
As the wind efficiency is reduced, a PISN mass range with the lower metallicity 
gets close to the zerometallicity estimate with no mass loss.
Provided the big uncertainty of the metallicity dependence,
the upper value of 1/3 $Z_\odot$ for the critical metallicity, 
below which the same PISN mass range as the range for the zerometallicity model is obtained, 
is suggested in \citet{langer+07}.
Meanwhile, the much higher value of $\gtrsim$300 $M_\odot$
are obtained for the lower end of the PISN initial mass range for
$\sim$1/5 $Z_\odot$ stars in other works \citep{yoshida&umeda11, yusof+13, yoshida+14}.
Apart from the wind mass loss, 
binary mass transfer can reduce the mass of the primary and increase the mass of the secondary stars.
Strong internal magnetic field may suppress the core convection 
to form a smaller mass CO core \citep{petermann+15},
while strong surface magnetic field may suppress the wind mass loss and thus help
to form a massive CO core on the contrary \citep{Georgy+17}.
Also a CO core may be extended by the rotation induced mixing 
\citep{Chatzopoulos&Wheeler12a, Yoon+12}.
All of those mechanisms can shift the initial mass-CO core mass relation
to affect the initial mass range for PISNe.

Most of the previous works have implicitly assumed that
the CO core mass range of PISNe is well-determined, and in fact,
less number of mechanisms have been suggested
to shift the CO core mass range for PISNe upward.
Multidimensional turbulence that may appear during core collapse and explosion will not affect
the hydrodynamical evolution, since the growth timescale is merely comparable to the 
timescale of the overall hydrodynamical evolution, and indeed,
multidimensional calculations have confirmed the PISN explosion \citep{chen+14a}.
A fast rotation of a CO core will regulate the collapsing motion
and may affect the CO core mass range for PISNe \citep[c.f.][]{chatzopoulos+13b, Chen15}.
However, a large specific angular momentum of $\gtrsim 10^{17}$ cm$^2$ s$^{-1}$
is required to affect the PISN explosion \citep{Glatzel+85}.
Considering the infrequency of the long gamma ray burst,
the progenitor of which is estimated to have a similar or smaller specific angular momentum of 
$\sim 10^{17}$ cm$^2$ s$^{-1}$ to fulfill the constraints of the collapsar model 
\citep{Woosley93, Yoon&Langer05, Woosley&Heger06},
it will be unreasonable to consider majority of VMSs
form fast rotating CO cores in the end.
Progenitor structures having an inflated envelope or not affect the explodability of the CO core,
because they have different temperature structures at the outer region of the core \citep{takahashi+18a} and 
a deflated envelope requires more momentum to be blown off \citep{kasen+11, takahashi+16}.
However, the change of the initial mass range by the effect is $\sim$ 10 $M_\odot$
and not so significant.

In this work, we investigate the evolution of VMSs
that form CO cores with various carbon-to-oxygen ratios,
and report the evolutionary consequences 
that significantly affect the CO core mass range for PISNe for the first time.
By introducing a multiplication factor $f_{\rm cag} \in [ 0.1, 1.2 ]$
to the $^{12}$C($\alpha, \gamma$)$^{16}$O reaction rate of \citet{Caughlan&Fowler88},
CO cores with $X(\rm{C})/X(\rm{O}) \sim$ 0.15--3.1 are developed.
In the next section,
firstly we provide the information of the stellar evolution code
and a short discussion on
the $^{12}$C($\alpha, \gamma$)$^{16}$O reaction rate,
then the results of VMS evolutions are discussed.
For more massive models, the later hydrodynamic evolution is
calculated by a 1D hydrodynamic code.
The code description is given in the first subsection in section 3.,
and the results are discussed in the second subsection.
Discussion of the event rate of PISNe and
of the observational consequences is presented in section 4.
Conclusion is given in the last section.

\section{Stellar Evolution Calculation}

\subsection{Method}

The stellar evolution of nonrotating zerometallicity VMSs is calculated
from zero age main sequence (ZAMS) until central carbon depletion at least or until iron core formation.
An initial mass of a model, $M_{\rm ini}$, is taken from [120, 460] $M_\odot$.
The initial composition is determined based on the result of Big bang nucleosynthesis of \citet{steigman07}.
Assuming all of the $^{2}$H has burned to form $^{3}$He before the ZAMS stage,
mass fractions of $^{1}$H, $^{3}$He, and $^{4}$He of 0.7599, $8.67\times10^{-5}$,
and 0.2400 respectively, are applied.

Calculations have been done by a stellar evolution code described in \citet{takahashi+16, takahashi+18a}, 
which was developed originally by Japanese researchers \citep{Saio+88a}.
In order to treat a general massive star evolution,
overall physical and numerical descriptions have been improved since then,
including the introduction of the wind mass loss and a large reaction network \citep{yoshida&umeda11},
the inertia term and the automatic mesh refinement scheme \citep{takahashi+13},
and stellar rotation \citep{takahashi+14}.
The equation of state in the code includes four species of particles, photon, ions, electron, and positron.
Photon is expressed as a blackbody radiation and
ions are approximated as the Boltzmann gas.
For the electron--positron part,
the reaction equilibrium between $e^-$$e^+$ pair creation and annihilation is assumed,
and analytic formulae for general Fermi-Dirac integrals are applied \citep{Blinnikov+96}. 
The convective overshooting is taken into account for the core hydrogen- and helium-burning phases.
An exponentially decaying function is applied with the overshoot parameter of $f_{\rm ov} = 0.015$,
with which a non-rotating solar metallicity models can account for the wide MS width 
observed for AB type stars in open clusters in our galaxy \citep{Maeder&Meynet89}.

A nuclear reaction network with 49 isotopes,
which includes all of the major nuclear reactions affecting the concerned evolution,
is incorporated in the stellar calculation.
The reaction rates are taken from the current version of JINA REACLIB \citep{Cyburt+10}
except for the $^{12}$C($\alpha, \gamma$)$^{16}$O reaction.

\subsection{The $^{12}$C($\alpha$,$\gamma$)$^{16}$O reaction rate}

Together with the 3$\alpha$ reaction,
the $^{12}$C($\alpha$,$\gamma$)$^{16}$O is an astrophysically important reaction,
which determines the $^{12}$C/$^{16}$O ratio in the universe.
In a He core of a massive star, the reaction takes place with a typical center-of-mass energy of 300 keV,
which results in a small cross section of $\sim2 \times 10^{-17}$ barn.
Since the cross section is far below the sensitivity of the current measurements,
experimental data that are obtained at the higher energy range
has to be extrapolated down to the astrophysically relevant energy range.
However, at this higher energy, the cross sections are complicated
by the interference from other excited states of $^{16}$O.
To disentangle the experimental data and to conduct a reliable extrapolation,
theoretical models such as the R-matrix theory \citep[e.g.][]{Azuma+10} are required.
Due to these complications in both experiment and theory,
the reaction rate has been still unsettled, 
despite numbers of investigations have been done over the years
\citep[c.f.][for a review]{deBoer+17}.

\begin{figure}[t]
	\centering
	\includegraphics[width=0.5\textwidth]{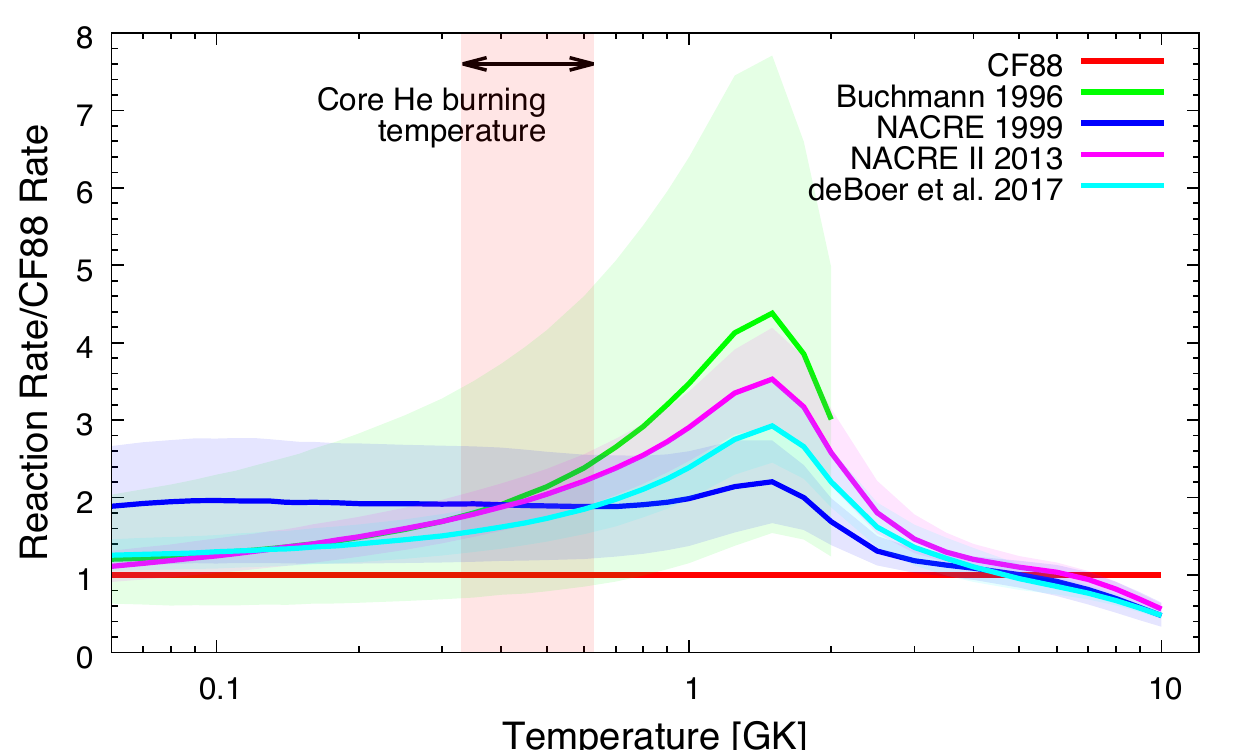}
	\caption{\footnotesize{Comparison of the $^{12}$C($\alpha$,$\gamma$)$^{16}$O reaction rate
	in the literature. The ratio between the rates of \citet{Buchmann96}, \citet{Angulo+99} (NACRE),
	\citet{Xu+13} (NACRE II), and \citet{deBoer+17} and the rate of \citet{Caughlan&Fowler88} (CF88)
	are shown with their uncertainties as functions of the temperature.
	The temperature range relevant for the core helium burning is also shaded by red.
	}}
	\label{fig-comparison}
\end{figure}
In Fig \ref{fig-comparison}, the $^{12}$C($\alpha$,$\gamma$)$^{16}$O reaction rates 
presented in the literature are compared.
In the relevant temperature range for core helium burning in a VMS of $2.1 \times 10^8$--$4.3 \times 10^8$ K,
a low value of $f_{\rm cag} \gtrsim 0.8$ is inside the uncertainty of the reaction rate of \citet{Buchmann96}.
On the other hand, the small reaction rate of \citet{Caughlan&Fowler88} is below the uncertainty ranges
of more recent works (\citealt{Angulo+99} (NACRE), \citealt{Xu+13} (NACRE II), \citealt{deBoer+17}).
Nevertheless, CO cores with various carbon-to-oxygen ratios are calculated
by applying a multiplication factor $f_{\rm cag} \in [0.1, 1.2]$ to the reaction rate of
$^{12}$C($\alpha,\gamma$)$^{16}$O of \citet{Caughlan&Fowler88} in this work.
This is because the aim of this work is to display a new route in the massive star evolution
that appears when the CO core has a high carbon-to-oxygen ratio.
In addition to the small reaction rate,
astrophysical origins such as additional mixing
may account for the large carbon-to-oxygen ratio.
Previous works have shown that
the carbon-to-oxygen ratio in a CO cores is influenced by 
the convective overshooting \citep{Imbriani+01}
and by the rotation induced mixing \citep{Chieffi&Limongi13}.

The $^{12}$C($\alpha,\gamma$)$^{16}$O reaction rate has also been inspected
by calculating theoretical yields of CCSNe
\citep{weaver&woosley93, timmes+95, woosley&heger07, West+13, Austin+17}.
Because carbon-to-oxygen ratio affects the electron mole fractions at the end of the stellar evolution,
the multiplier of the reaction rate has correlation and anti-correlation to even-Z and odd-Z elemental yields.
The same trends have also been found in our calculations of 15 $M_\odot$ models,
having the converging point at $f_{\rm cag} = 1.2$.
Thus the upper value of $f_{\rm cag} = 1.2$ is the one used in our conventional calculations.
Since the resulting carbon-to-oxygen ratio of $\sim 0.15$ is small,
we expect that the evolution applying $f_{\rm cag} = 1.2$ is qualitatively similar to the one from pure oxygen cores.
For the same reason, calculations with $f_{\rm cag} = 1.2$ will represent
calculations with higher $f_{\rm cag} > 1.2$.
This is why we drop calculations with $f_{\rm cag} > 1.2$ from this work,
despite the larger reaction rate is more compatible with recent estimates.

\subsection{Result}

\begin{figure}[t]
	\centering
	\includegraphics[width=0.5\textwidth]{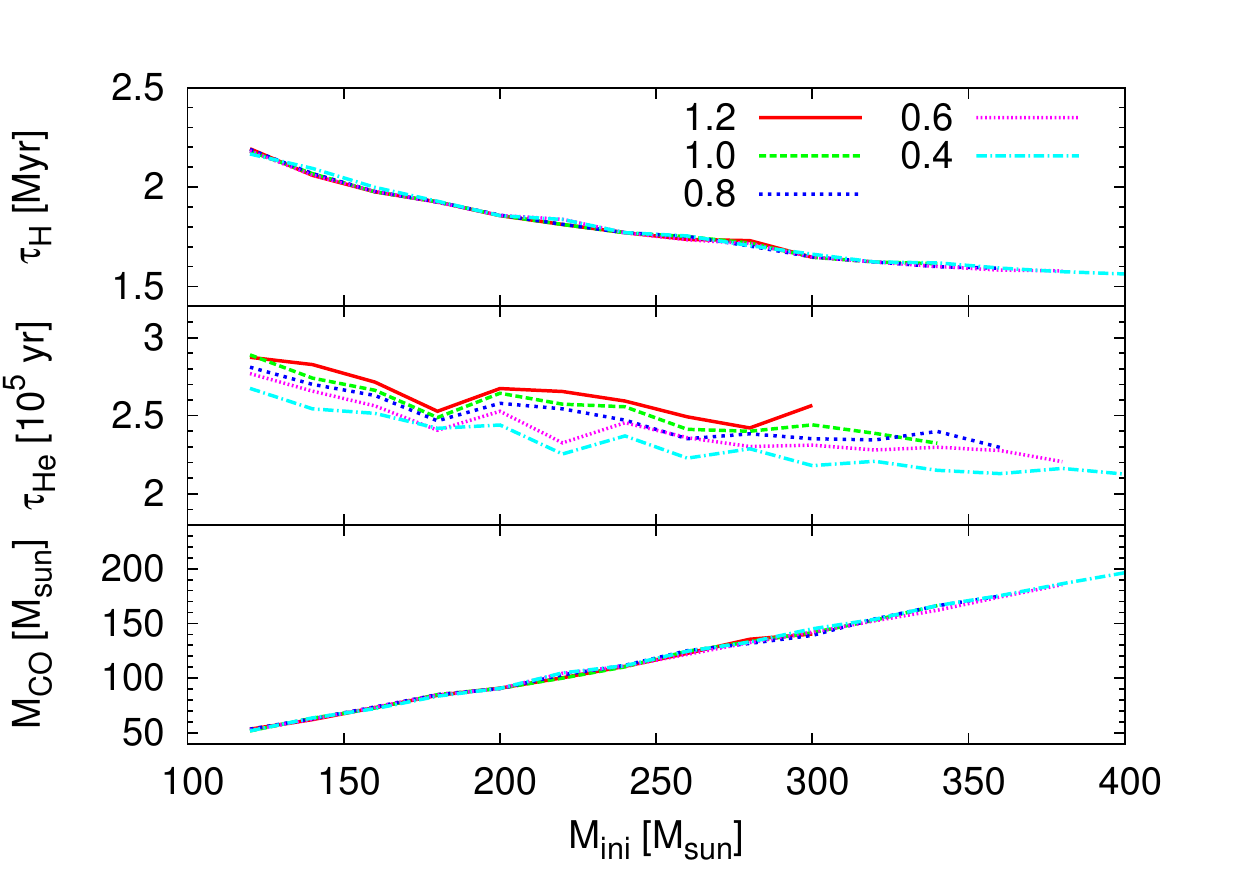}
	\caption{\footnotesize{Durations of core hydrogen- (top) and helium-burning (middle) phases
	 and CO core masses (bottom) are shown for models with
	 $f_{\rm cag}$ = 1.2 (red, solid), 1.0 (green, long dashed), 0.8 (blue, short dashed),
	 0.6 (magenta, dotted), and 0.4 (cyan, dash-dotted).}}
	\label{fig-summary}
\end{figure}

Until core carbon depletion, a zerometallicity VMS experiences
core hydrogen- and helium-burning phases.
In Fig.~\ref{fig-summary}, durations of these phases, $\tau_{\rm H}$ and $\tau_{\rm He}$,
and resulting CO core masses, $M_{\rm CO}$, which is defined as
the innermost mass coordinate where helium mass fraction exceeds $X({\rm He}) > 10^{-2}$,
are shown as functions of the initial mass, $M_{\rm ini}$,
for selected sequences of $f_{\rm cag}$.
The duration of the hydrogen-burning phase is independent\footnote{Actually, 
tiny differences in $\tau_{\rm H}$ are seen for models with the same initial masses.
However, these differences should not have a physical significance
but have a numerical origin.
The different $f_{\rm cag}$ affects the evolution of a zerometallicity VMS 
during the pre-ZAMS He burning phase.
This difference is enhanced through the evolution of the convective regions around the H burning core,
because the convective evolution is significantly sensitive to any kind of numerical errors.
Finally, the merging of shell- and central convections takes place
during the core hydrogen-burning phase in some models,
causing the different $\tau_{\rm H}$.}

of $f_{\rm cag}$ but depends on the initial mass, 
because the $^{12}$C($\alpha$,$\gamma$)$^{16}$O reaction is irrelevant to the hydrogen burning.
On the other hand, models with smaller $f_{\rm cag}$
tend to have slightly shorter helium burning phases
for models with the same initial masses.
The CO core mass is again independent from $f_{\rm cag}$.
This is because the size of the convective core is nearly constant
during the helium-burning phase.

\begin{figure*}[t]
	\centering
	\includegraphics[width=0.9\textwidth]{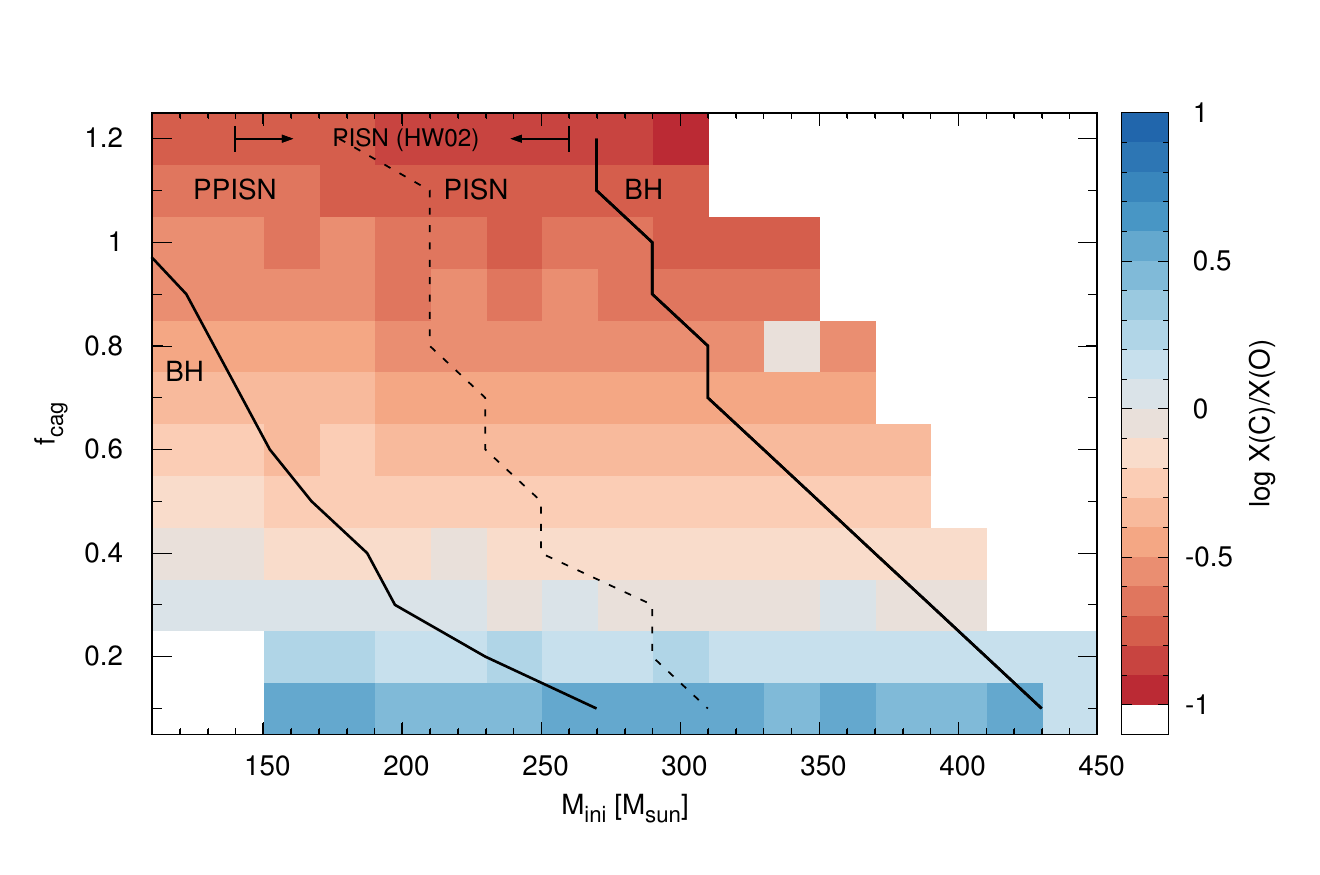}
	\caption{\footnotesize{Phase diagram of the zerometallicity VMSs.
	The color shows the central $X({\rm C})/X({\rm O})$.}}
	\label{fig-phase}
\end{figure*}

Hence, the most relevant consequence from applying different $f_{\rm cag}$ is
the different carbon-to-oxygen ratio in the same mass CO core.
The central $X({\rm C})/X({\rm O})$ measured
when the central temperature reaches $\log T_c {\rm [K]}= 8.8$
is shown by a color map in Fig.\ref{fig-phase}.
The phase space is divided into 4 regions
according to the fate of the model,
while the definition of each boundary is explained later.
Also, the initial mass range of PISNe indicated by \citet{heger&woosley02} is shown.
The color map shows that 
models with small $f_{\rm cag}$ have high $X({\rm C})/X({\rm O})$.
More massive models tend to have less $X({\rm C})/X({\rm O})$, however,
the mass dependency is much weaker than the $f_{\rm cag}$ dependency.
Thus, models with $f_{\rm cag} = 1.2$ have the lowest $X({\rm C})/X({\rm O}) \sim 0.15$,
models with the intermediate $f_{\rm cag} = 0.6$ have $X({\rm C})/X({\rm O}) \sim 0.46$,
and models with $f_{\rm cag} = 0.1$ have the highest $X({\rm C})/X({\rm O}) \sim 3.1$.

\begin{figure*}[t]
	\centering
	\includegraphics[width=0.85\textwidth]{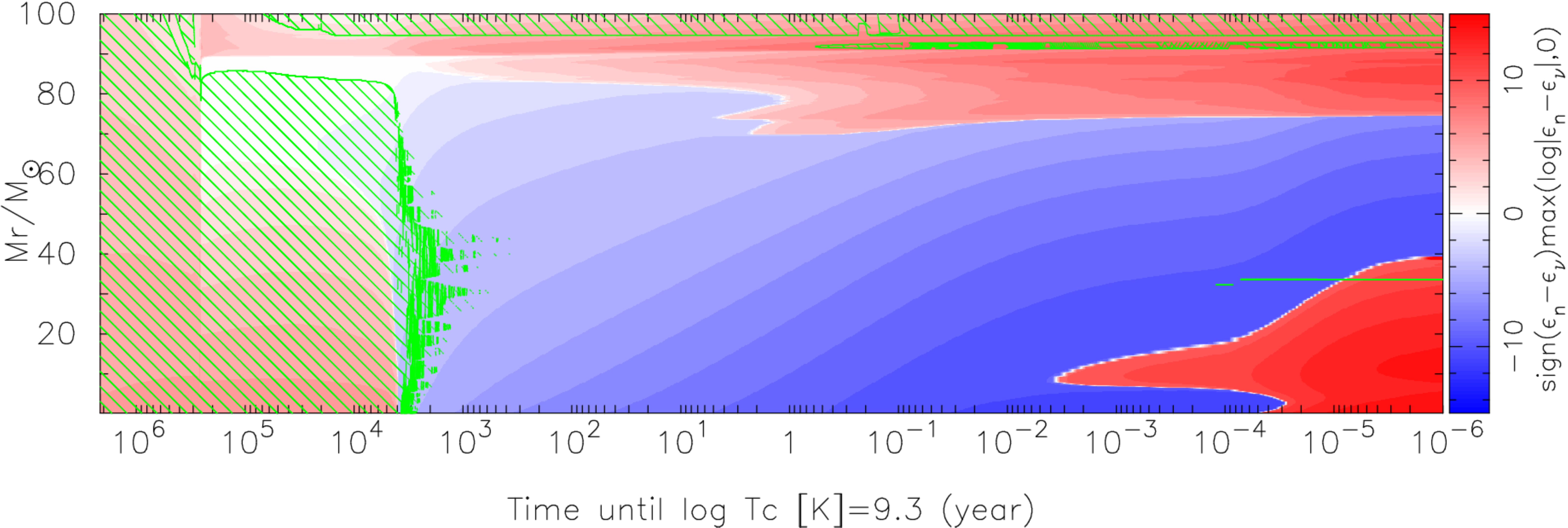}
	\includegraphics[width=0.85\textwidth]{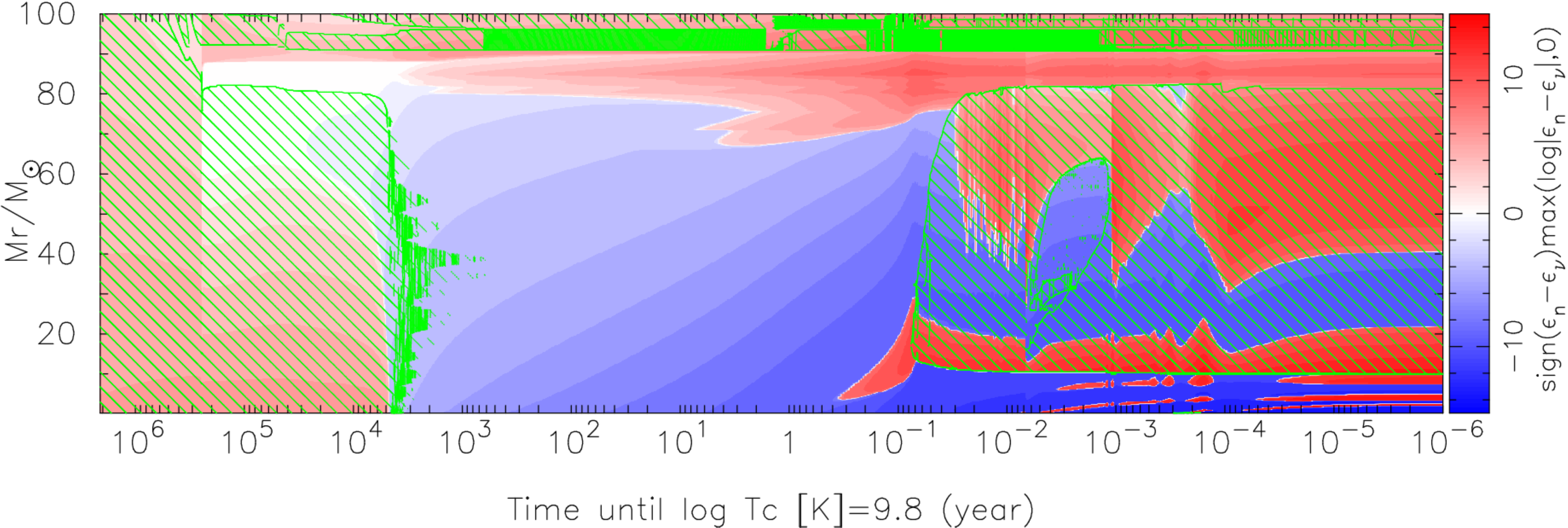}
	\caption{\footnotesize{Kippenhahn diagrams of 180 $M_\odot$ models
	with $f_{\rm cag} =$ 1.2 (top) and 0.3 (bottom).
	The convective evolution is shown from the ZAMS phase until log $T_c$ [K]$=9.3$
	for the model with $f_{\rm cag} =$ 1.2, while that until log $T_c$ [K]$=9.8$
	is shown for the model with $f_{\rm cag} =$ 0.3.
	Green-hatched regions show the convective regions.
	Colors indicate the net heating (red) or the cooling (blue) rates at the region.}}
	\label{fig-khd-cconv}
\end{figure*}

We have found that, due to the high core carbon fraction,
less massive models with small $f_{\rm cag}$ develop
shell convection during the core carbon-burning phase.
The thick solid line passing from $M_{\rm ini} =$ 110 $M_\odot$ at $f_{\rm cag} = 1.0$
to $M_{\rm ini} =$ 270 $M_\odot$ at $f_{\rm cag} = 0.1$ in Fig. \ref{fig-phase}
is the upper boundary of models that experience this convective shell formation.
Figure \ref{fig-khd-cconv} shows the evolution of convective regions
for models of $M_{\rm ini} = 180$ $M_\odot$ 
with different $f_{\rm cag} =$ 1.2 (top panel) and 0.3 (bottom panel).
No convection develops for the $f_{\rm cag} = 1.2$ case,
which has a small $X({\rm C})/X({\rm O})$ of 0.18.
On the other hand, a large shell convective region appears at $M_r \gtrsim 10$ $M_\odot$
from $\sim 7 \times 10^{-2}$ yr before the calculation end for the $f_{\rm cag} = 0.3$ case,
which has 5.7 times larger ratio of $X({\rm C})/X({\rm O}) =$ 1.04.

\begin{figure}[t]
	\centering
	\includegraphics[width=0.5\textwidth]{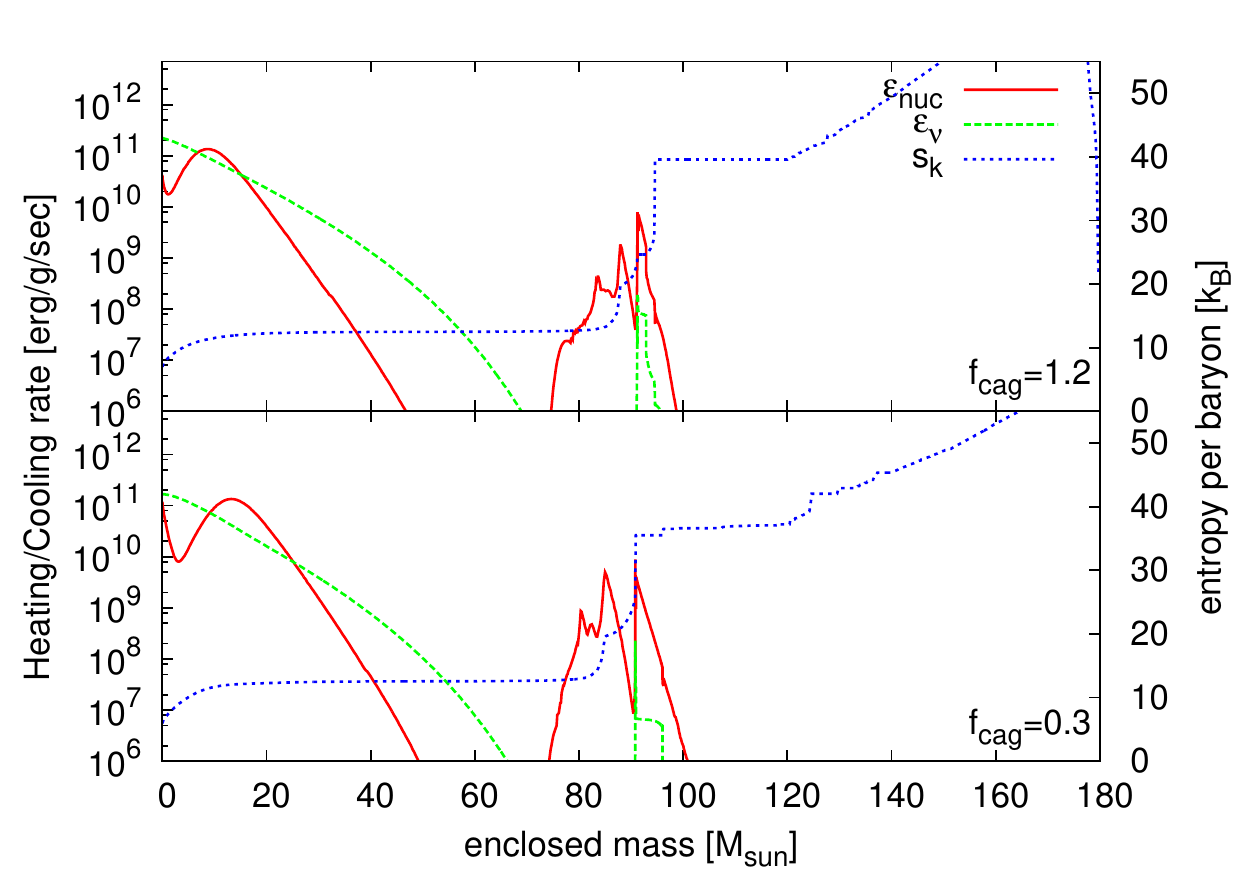}
	\caption{\footnotesize{Distributions of nuclear heating rate (red, solid),
	neutrino cooling rate (green, dashed), and entropy per baryon (blue, dotted)
	are shown for $M_{\rm ini} = 180$ $M_\odot$ models with
	$f_{\rm cag} =$ 1.2 (top) and 0.3 (bottom).
	Both models have the same central temperatures of $\log T_c {\rm [K]}= 9.18$.}}
	\label{fig-edist}
\end{figure}

In general, a CO core material easily becomes convectively unstable
if a certain amount of heating takes place in the shell region.
This is because a newly-formed CO core in a VMS
has nearly homogeneous distributions of entropy and chemical composition
as a result of the effective mixing during the previous core helium-burning phase.
In addition, neutrino cooling, which triggers the further core evolution by reducing the core entropy,
is dominated to occur in the central region of the core.
Therefore, the isentropic structure in the surrounding region remains.
Figure \ref{fig-edist} shows distributions of heating and cooling rates and the entropy for the two models
when the central temperatures become $\log T_c {\rm [K]}= 9.18$.
The figure shows that the $f_{\rm cag} = 0.3$ model has
a shell region at $M_r \sim 10$--$25$ $M_\odot$
where the nuclear heating rate significantly exceeds the neutrino cooling rate.
This net heating soon creates a negative entropy gradient
and drives shell convection at that region.
On the other hand, the nuclear heating rate is only slightly larger than
the neutrino cooling rate in the $f_{\rm cag} = 1.2$ model.
Except for the carbon depleted central region, 
the heatings are solely caused by the $^{12}\rm{C} + ^{12}\rm{C}$ reaction.
Therefore, the heating rate is proportional to the the square of the carbon mass fraction.
Given the similar core entropies the lower heating rate
is explained by the lower core carbon fraction.
Finally no convection appears in this model until core collapse sets in.

\begin{figure}[t]
	\centering
	\includegraphics[width=0.5\textwidth]{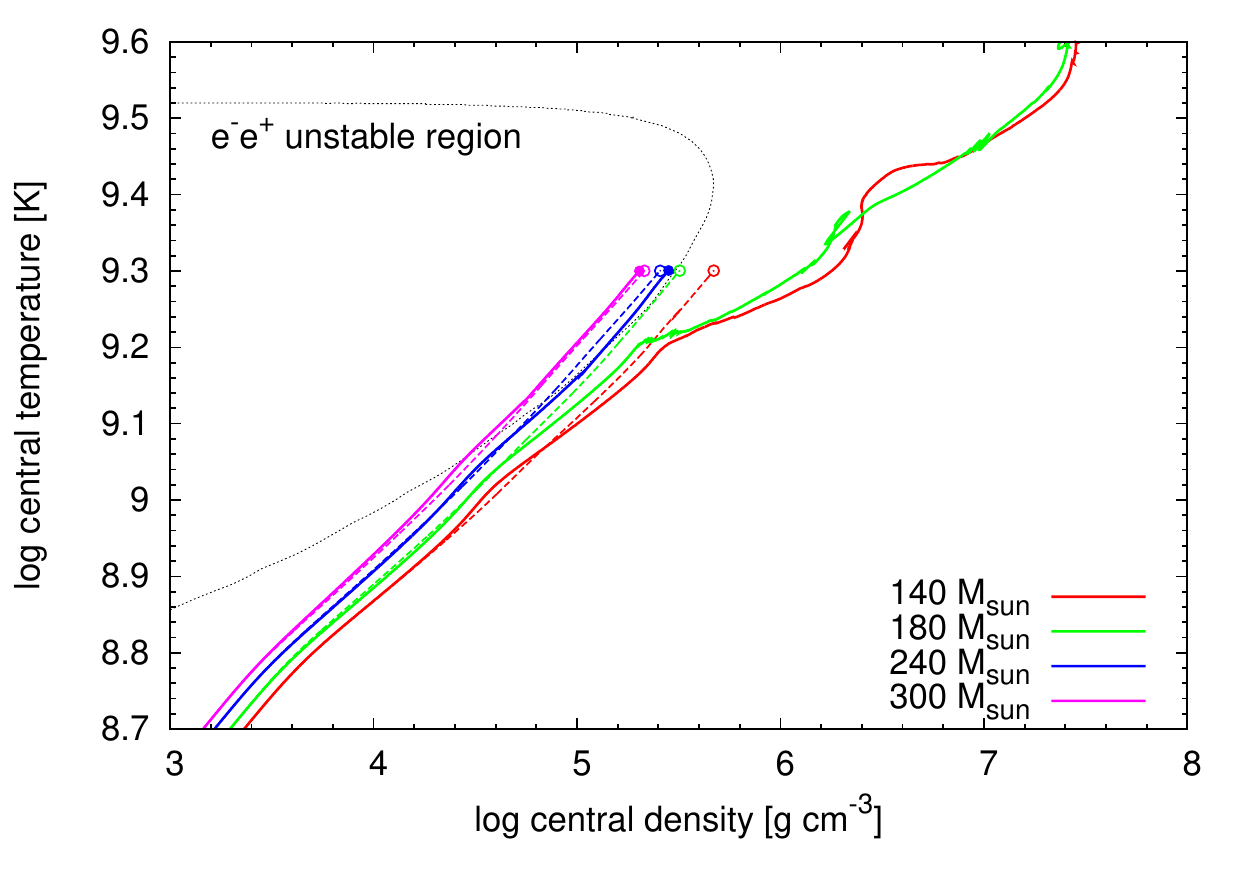}
	\caption{\footnotesize{The central density and temperature evolution.
	Models with $f_{\rm cag} = 0.3$ are shown by solid lines
	while models with $f_{\rm cag} = 1.2$ are by dashed lines.
	Selected initial masses are 140 (red), 180 (green), 240 (blue), and 300 (magenta) $M_\odot$.
	The boundary of the pair instability region is shown by the black dotted line.
	}}
	\label{fig-rhot}
\end{figure}

The evolution of central density and temperature is shown in Fig.~\ref{fig-rhot} for selected models.
All of the models with $f_{\rm cag} = 1.2$, shown by dashed lines, are nonconvective.
For models with $f_{\rm cag} = 0.3$ shown by solid lines,
less massive models with $M_{\rm ini} \leq 195$ $M_\odot$ develop
shell convection during the core carbon-burning phase,
while more massive models are nonconvective.
It is evident that those convective models offset to the lower entropy side,
i.e., the higher density for the same temperature,
when the central temperatures reach $\log T_c {\rm [K]} \sim 9.2$.
The branching moments are exactly when
the shell convections develop in the cores.
Thus, due to the emergence of the shell convective regions,
the effective core masses of the convective models are reduced.
And due to the effective neutrino cooling,
the central entropies rapidly decrease to match with the new core masses.
As a consequence, the low entropy core avoids 
being affected by the $e^-$$e^+$ pair-creation instability.
No dynamical collapse or energetic pulsations take place for the convective models.
The two convective models shown in the Fig.~\ref{fig-rhot} form hydrostatic iron cores in the end.

Based on the result, we estimate that the formation of a stellar mass black hole (BH),
instead of the explosion as a PISN, is the fate of convective models. 
Therefore, the minimum mass to be affected by
the $e^- e^+$ pair creation instability is significantly shifted upward,
for example, $M_{\rm ini} > 160$ $M_\odot$ for models with $f_{\rm cag} = 0.6$ and $X$(C)/$X$(O) $\sim 0.46$
and $M_{\rm ini} > 280$ $M_\odot$ for models with $f_{\rm cag} = 0.1$ and $X$(C)/$X$(O) $\sim 3.1$.
Note that weak pulsations possibly appear in the outer region of the CO core in reality,
which are dumped in a time-implicit evolutionary calculation with a long time step.
Although the nonlinear coupling with carbon burning may trigger mass ejection
and further affect the evolution, this is beyond the scope of this work.

\section{Hydrodynamic Calculation}

\subsection{Method}

After the central carbon depletion, nonconvective models 
with more massive initial masses of $\gtrsim140$ $M_\odot$ are affected
by the $e^-$$e^+$ pair creation instability.
The late hydrodynamic evolution is calculated
by a general relativistic hydrodynamic code described in \citet{yamada97}.
A result at log $T_c$ [K] $\sim9.2$ calculated by the stellar evolution calculation is used for the initial structure.
Except for the small reaction network with reduced 49 isotopes, 
which is identical to the evolution calculation in this work,
the code settings are the same as in \citet{takahashi+16, takahashi+18a}.
The equation of state, the local neutrino cooling rate, and the reaction network
are imported from the stellar evolution code \citep{takahashi+16}.

Chemical mixing and energy transport by convection are not considered in the hydrodynamic code.
The code thus has no capability to model the Rayleigh-Taylor (RT) instability 
developing at the core-envelope interface associated with the reverse shock, 
which may affect the efficiency of the fallback and thus 
the bounding mass between pulsationall-PISNe and PISNe. 
However, no major instabilities will be developed during the neon and oxygen burning phases, 
and therefore the explosion of PISNe will not be affected by the omission of convection.
\citet{chen+14a} has shown that only a mild instability
grows by the oxygen burning in their 2D simulations. 
This can be understood as the growth time, 
which may be estimated as $t_{\rm growth} \sim |N^2|^{-1/2}$, 
where $N$ is the Brunt-V\"{a}is\"{a}l\"{a} frequency, is $\sim10$ sec at its minimum and 
is merely comparable to the timescale of core contraction and expansion around the turning point. 
Soon after the instability starts to grow, the core expands and the growth time will be significantly lengthened. 
Thus the instability is expected to freeze out for the further evolution, which is what observed in the 2D simulations.

\subsection{Result}

\begin{figure}[t]
	\centering
	\includegraphics[width=0.5\textwidth]{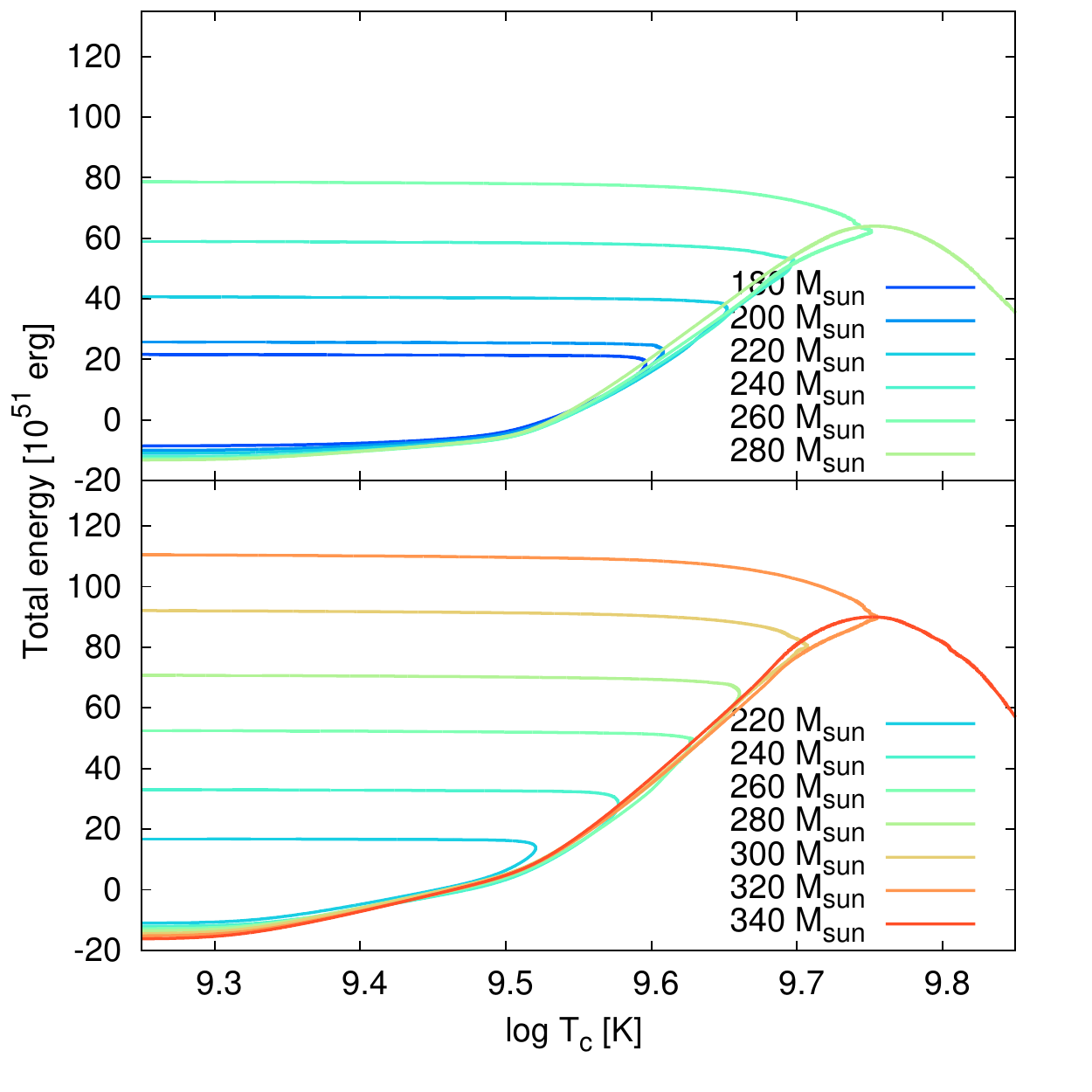}
	\caption{\footnotesize{The evolution of the total energy during explosion.
	Results of models with $f_{\rm cag} =$ 1.2 and 0.6 are respectively shown
	by the top and the bottom panels.}}
	\label{fig-teng}
\end{figure}

In Fig.~\ref{fig-teng}, the evolution of the total energy, which is defined as
$E_{\rm tot} = \int (\frac{1}{2}U^2 + e^{\rm therm+pair} - \frac{GM_b}{r})dM_b$,
is shown as a function of the central temperature.
Here, $\rho_b e^{\rm therm+pair} \equiv \rho_b e^{\rm therm} + \rho_{\rm pair} c^2$
is the internal energy density including the rest mass of created electron-positron pair
\citep[for detailed information, see][]{takahashi+16} and
$U$ and $r$ are the velocity and the radius of the mass shell
at the enclosed baryon mass of $M_b$.

Results of models with $f_{\rm cag} = 1.2$ shown in the top panel
are essentially the same as reported in \citet{takahashi+16}.
During the contraction, the total energy firstly increases due to the neon burning
that initiates when the central temperature reaches log $T_c$ [K] $\sim$ 9.3.
However, the released energy is small and the star k-eps-converted-to.pdf contracting.
Next, oxygen burning sets in after the central temperature reaches log $T_c$ [K] $\sim$ 9.5,
significantly increasing the total energy.
The star returns its contracting motion to expansion
when the large enough energy is released by the reaction.
On the other hand, when the released energy is insufficient and
the central temperature reaches log $T_c$ [K] $\sim$ 9.75,
the next important reaction of the photodisintegration initiates.
Because this reaction converts the internal energy into rest mass of nuclei,
the total energy defined above rapidly decreases.
Empirical results obtained by \citet{takahashi+16} show that
the star finally collapses to form a BH if
the central temperature exceeds log $T_c$ [K] $\sim$ 9.8.

The second panel shows the total energy evolution of models with $f_{\rm cag} = 0.6$.
The basic picture discussed for models with $f_{\rm cag} = 1.2$
is still applicable to other cases with $f_{\rm cag} < 1.2$.
I.e., important temperatures of $T_c$ [K] $\sim$ 9.3, 9.5, and 9.8
divide the hydrodynamic evolution into four phases.
On the other hand, the figure also shows that 
the inclination of energy increase during the neon-burning phase
becomes steeper than the case of $f_{\rm cag} = 1.2$.
This is because a model with $f_{\rm cag} = 0.6$ has a higher core neon fraction.
Neon is the prime product of the carbon burning of
\begin{eqnarray}
	^{12}\rm{C} + ^{12}\rm{C} &\rightarrow& ^{20}\rm{Ne} + ^{4}\rm{He} \nonumber \\
	^{16}\rm{O} + ^{4}\rm{He} &\rightarrow& ^{20}\rm{Ne}, \nonumber
\end{eqnarray}
so that the high core carbon fraction before the core carbon-burning phase
results in the high core neon fraction after the core carbon depletion.
Moreover, the neon burning in the surrounding region
has a major contribution for the energy increase
even after the central temperature reaches log $T_c$ [K] $\sim$ 9.5.
Therefore, a low $f_{\rm cag}$ model with the same initial mass
has a larger total energy for the same central temperature.
As a consequence, a model with lower $f_{\rm cag}$
returns its contracting motion to explosion
having a lower central temperature.
In other words, models with lower $f_{\rm cag}$ are more easily
explode than models with higher $f_{\rm cag}$.

\begin{figure*}[t]
	\centering
	\includegraphics[width=\textwidth]{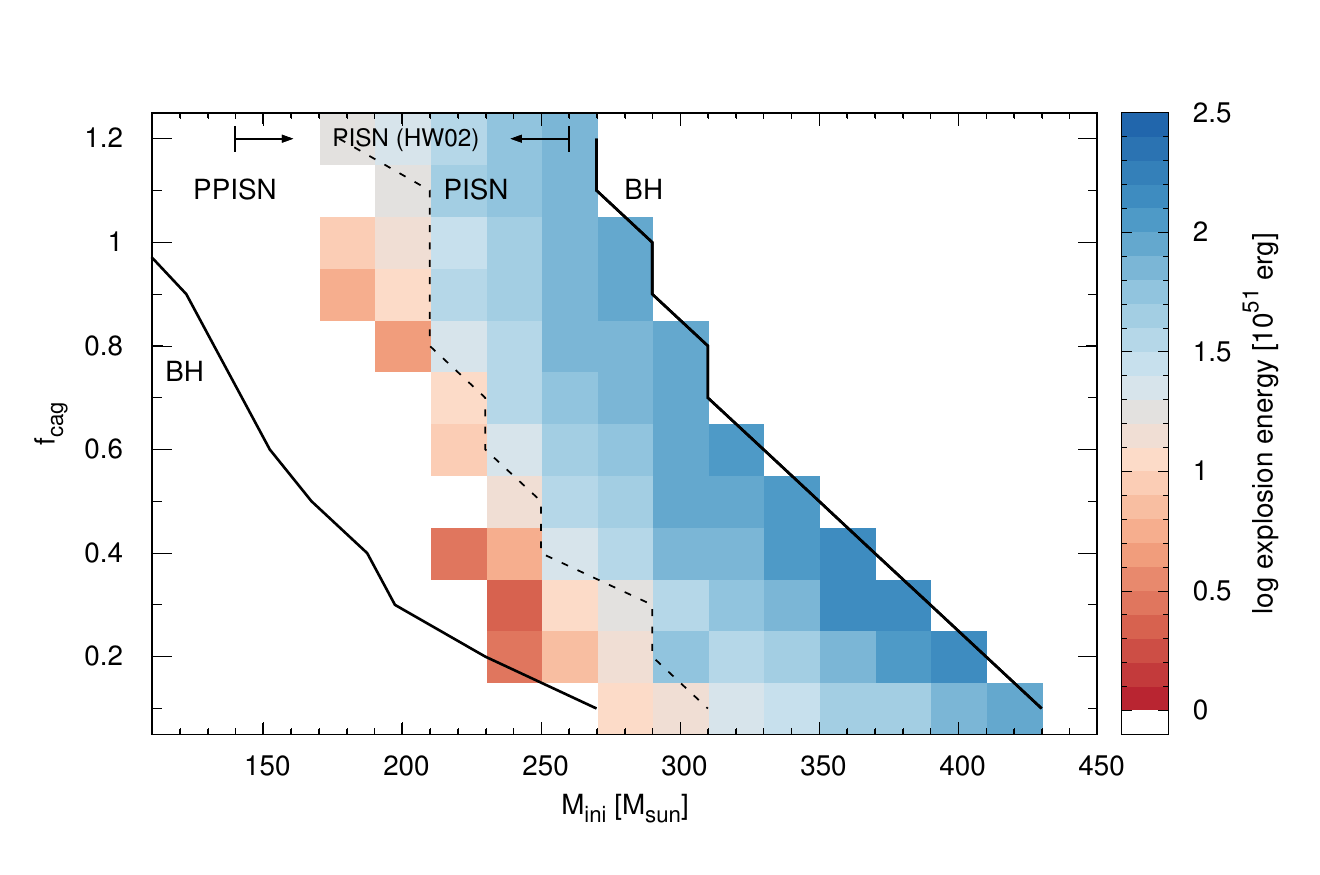}
	\caption{\footnotesize{The phase diagram of the zerometallicity VMSs.
	For models that have positive total energies after the core collapse,
	the energies $\sim 10^4$ sec after the core collapse are shown by the colors.
	}}
	\label{fig-eeng}
\end{figure*}

The phase diagram of the fate is again shown in Fig.~ \ref{fig-eeng}
but with the color map showing the total energy
after $10^4$ s from the start of the hydrodynamic calculation.
Note that the total energies are shown only for models
that have positive total energies after the core collapse.
The thick solid line passing from $M_{\rm ini} =$ 270 $M_\odot$ at $f_{\rm cag} = 1.2$
to $M_{\rm ini} =$ 430 $M_\odot$ at $f_{\rm cag} = 0.1$ shows a boundary
between progenitors of PISNe and BH formation.
The definition is clear: whether the model returns the contracting motion or not.

The boundary between progenitors of pulsational PISNe (PPISNe) and PISNe is shown
by a\sout{nother} dashed line passing from $M_{\rm ini} =$ 175 $M_\odot$ at $f_{\rm cag} = 1.2$
to $M_{\rm ini} =$ 310 $M_\odot$ at $f_{\rm cag} = 0.1$.
Similar to PISNe, PPISNe are triggered by the $e^- e^+$ pair creation instability.
However, in this case, a central part of the star remains gravitationally bound after the expansion
because of the smaller energy injection by the thermonuclear reactions
\citep{woosley+07, chatzopoulos&wheeler12b, chen+14a, yoshida+16a, woosley17}.
In this work, an expanding model is considered as a PPISN 
if the central mesh of the model restarts contraction after its first expansion
during $10^4$ s from the start of the hydrodynamic calculation.
The bounding mass at $f_{\rm cag}=1.2$ is larger than the results of \citet{heger&woosley02},
however, this discrepancy will be well explained as
hydrogen-rich envelopes are included in our calculation \citep{kasen+11, takahashi+16}.
The central remnant of a PPISN is considered to restart
hydrostatic evolution leading to the iron core collapse in the end.
Because of the high masses, the fate of PPISN models are determined as BH formation.

The phase diagram clearly shows the strong dependence
of the initial mass range of PISNe on the core carbon-to-oxygen ratio.
With the highest $f_{\rm cag} = 1.2$, the models have small $X({\rm C})/X({\rm O}) \sim 0.15$
and have the lower shifted initial mass range of $M_{\rm ini} \in [ 175, 270 ]$ $M_\odot$.
The initial mass range becomes $M_{\rm ini} \in [ 240, 320 ]$ $M_\odot$
for models with the intermediate $f_{\rm cag} = 0.6$ with $X({\rm C})/X({\rm O}) \sim 0.46$,
and the highest shifted mass range of $M_{\rm ini} \in [ 310, 430 ]$ $M_\odot$ results from
models with the lowest $f_{\rm cag} = 0.1$ which have the largest $X({\rm C})/X({\rm O}) \sim 3.1$.

\begin{figure}[t]
	\centering
	\includegraphics[width=0.5\textwidth]{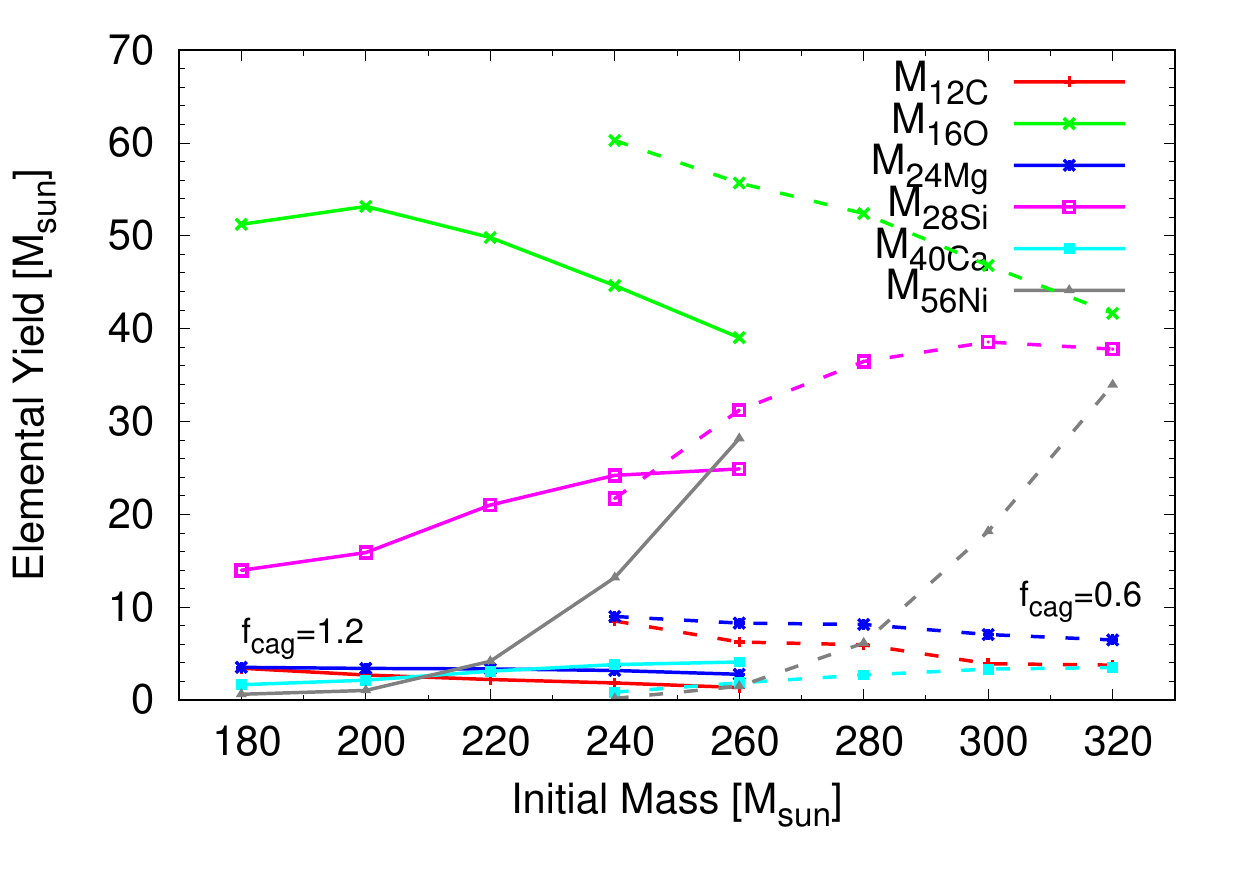}
	\caption{\footnotesize{The representative elemental yields
	($^{12}$C by red,
	$^{16}$O by green,
	$^{24}$Mg by blue,
	$^{28}$Si by magenta,
	$^{40}$Ca by cyan, and
	$^{56}$Ni by gray)
	ejected by a PISN are shown.
	Results of models with $f_{\rm cag} = 1.2$ and 0.6 are
	shown by solid and dashed lines, respectively.
	}}
	\label{fig-yield}
\end{figure}

In spite of the very different initial mass ranges,
the initial mass dependences of the explosion energy as well as the elemental yields
are quite similar for models with different $f_{\rm cag}$.
The smallest explosion energy k-eps-converted-to.pdf $\sim 10$--$30 \times 10^{51}$ erg for a wide range of $f_{\rm cag}$,
although the initial mass spans a wide range of $M_{\rm ini} \in [ 175, 310 ]$ $M_\odot$.
And the largest explosion energy increases from $\sim 80 \times 10^{51}$ erg
to $\sim 130 \times 10^{51}$ erg, while the initial mass increases from 260 $M_\odot$
for the $f_{\rm cag} = 1.2$ models to 400 $M_\odot$ for the $f_{\rm cag} = 0.2$ models.
The yields of representative elements are shown in Fig. \ref{fig-yield}
for exploding models with $f_{\rm cag} = 1.2$ and 0.6.
It clearly shows the similar mass dependencies.
As a result, models with the same mass but with different $f_{\rm cag}$ 
can produce totally different explosion energy and elemental yields.
For example, 260 $M_\odot$ models have quite different $^{56}$Ni yields of
1.48 $M_\odot$ for the model with $f_{\rm cag} = 0.6$ and 28.1 $M_\odot$ for the model with $f_{\rm cag} = 1.2$.

\section{Discussion}

\subsection{PISN event rate}

In order to demonstrate the impact of a larger mass range of PISN progenitors,
we estimate the event rate of PISNe by applying a simple initial mass function (IMF)
that is characterized by the slope ($\alpha$) and the upper limiting mass ($M_{\rm up}$) above which no star is born.
A typical $\alpha$ for low redshift PISNe will be the Salpeter value, $\alpha$ = 2.35
(however, a small IMF slope of $\alpha \sim 1.90$ is
recently obtained for a massive stellar cluster R136; \citealt{Schneider+18}).
A flat value $\alpha$ = 0 may be applicable for the IMF of zerometallicity stars \citep[c.f.][]{hirano+14}.
For $M_{\rm up}$, while a low value of $M_{\rm up} < 200$ $M_\odot$ for finite metallicity stars
has been rejected by the population synthesis for the cluster R136 \citep{Schneider+14},
the actual value is quite uncertain for both the low-redshift universe and the early universe
(for example, the maximum mass for zero-metallicity stars is estimated as
$\lesssim$ 300 $M_\odot$ in \citet{susa+14} and
$\lesssim$ 1000 $M_\odot$ in \citet{hirano+15}).
Therefore, massive values of $> 200$ $M_\odot$ are tested here as a free parameter.

Instead of the absolute value, a relative event rate of PISNe to a rate of CCSNe,
\begin{eqnarray}
	\lambda(\alpha, M_{\rm up}, f_{\rm cag})
		= \frac{ \int^{{\rm min}(M_{\rm PISN, max}, M_{\rm up})}_{{\rm min}(M_{\rm PISN, min}, M_{\rm up})}
				M^{\rm -\alpha} dM_{\rm ini} }
			{ \int^{20 M_\odot}_{10 M_\odot} M^{\rm -\alpha} dM_{\rm ini} }
\end{eqnarray}
is calculated, in which the mass range of 10 to 20 $M_\odot$ is assumed for CCSN progenitors.
The relative rate $\lambda$ depends on $f_{\rm cag}$ through $M_{\rm PISN, min}$ and $M_{\rm PISN, max}$,
which are the minimum and the maximum masses for PISN progenitors, respectively.

Here we assume that the initial mass range for PISN progenitors only depends on $f_{\rm cag}$
and is independent especially from the metallicity of the star.
This can be justified for low metallicity environments with $Z < 1/3 Z_\odot$ \citep{langer+07}, 
since a CO core formed in a star with the same initial mass has a nearly metallicity-independent mass 
unless the efficient wind mass loss significantly reduces the mass of the star.
We does not take the effect of the envelope structure into account in this estimate.
Whether the envelope of a VMS inflates or not affects the explodability
and thus changes the initial mass range of the PISNe \citep[e.g.,][]{takahashi+18a}.
However, the shift is $\sim 10$ $M_\odot$ and
much less effective than the effect of core carbon fraction considered here.

Also we does not take the effect of the rotational induced mixing into consideration.
This is because the efficiency of the rotational mixing is highly uncertain.
Fast rotating VMSs in \citet{Chatzopoulos&Wheeler12a} form extended He cores,
as a result, the lower and higher end of the PISN mass range shifts to lower masses.
For example, their 95 $M_\odot$ model with 30\% $v_{\rm ZAMS}/\sqrt{1-\Gamma}v_{\rm Kep}$,
where $v_{\rm ZAMS}$ is the surface rotation velocity at the ZAMS phase,
$v_{\rm Kep} = \sqrt{GM/R}$ is the Kepler velocity, and 
$\Gamma = L/L_{\rm Edd}$ is the Eddington factor, forms a 90 $M_\odot$ oxygen core, 
which has 40 $M_\odot$ higher mass than the nonrotating counterpart.
On the other hand, a 100 $M_\odot$ model with 47\% $v_{\rm ZAMS}/\sqrt{1-\Gamma}v_{\rm Kep}$
which also has $v_{\rm ZAMS} = 704$ km s$^{-1}$ and 30\% $v_{\rm ZAMS}/v_{\rm Kep}$
in \citet{Yoon+12} develops a 65.81 $M_\odot$ CO core,
and a 85 $M_\odot$ model with $v_{\rm ZAMS} = 800$ km s$^{-1}$ 
in \citet{Ekstroem+08b} forms a 43.92 $M_\odot$ CO core.
They are merely 13.88 and 9.42 $M_\odot$ larger than 
their nonrotating counterparts, respectively.
Also in \citet{takahashi+18a}, less effective enhancements of 5.80 and 2.29$M_\odot$ 
are obtained in their 100 $M_\odot$ rotating models of 30\% $v_{\rm ZAMS}/v_{\rm Kep}$ 
with and without the Tayler-Spruit dynamo.

\begin{figure}[t]
	\centering
	\includegraphics[width=0.5\textwidth]{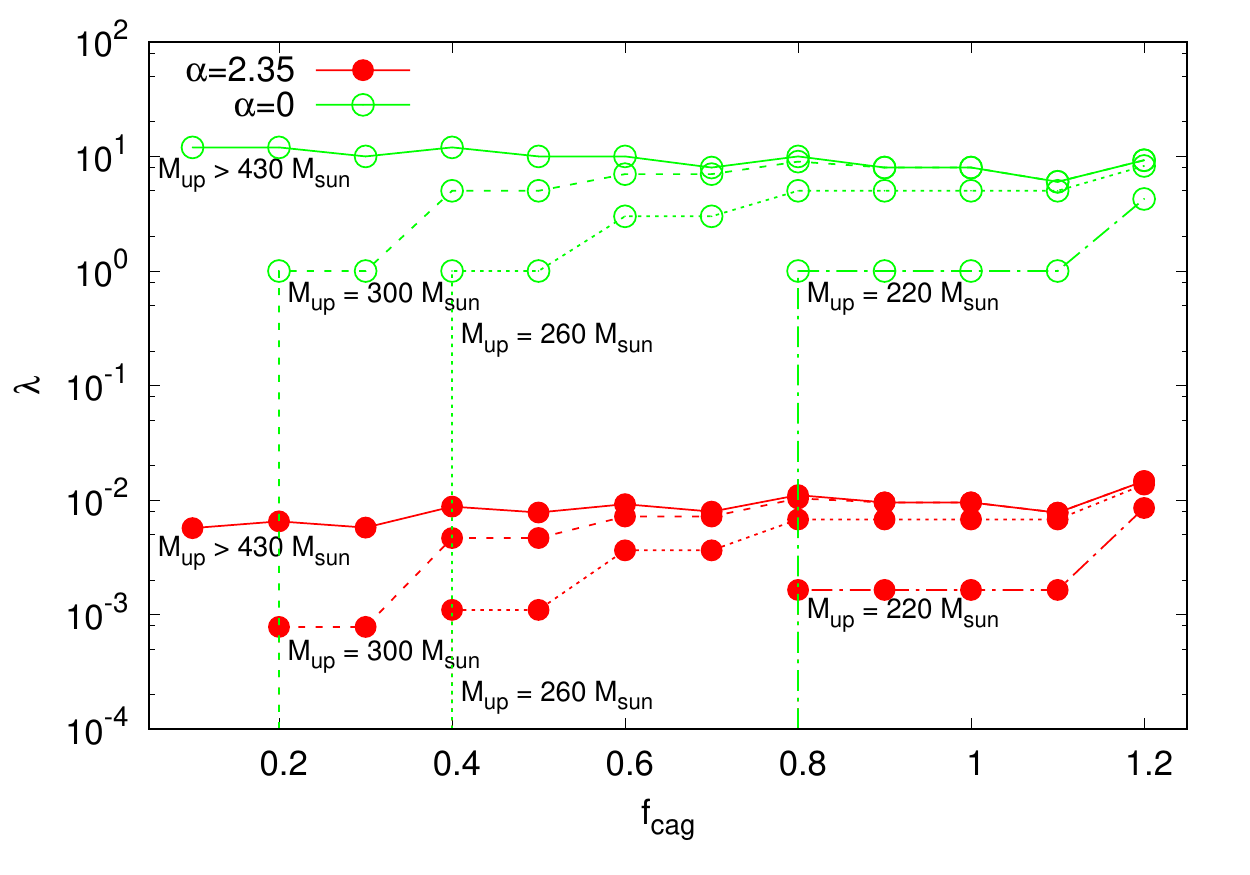}
	\caption{\footnotesize{Relative event rate of PISNe.
	Different colors indicate results with different IMF indexes of
	$\alpha = 2.35$ by red and $\alpha = 0$ by green.
	Results with considerably large $M_{\rm up} > 430 M_\odot$ are shown by solid lines,
	and results with $M_{\rm up} = 300, 260, 220 M_\odot$ are shown by
	dashed, dotted, and dash-dotted lines, respectively.}}
	\label{fig-fraction}
\end{figure}
Results are shown in Fig.~\ref{fig-fraction}.
Similar properties are deduced for both results of $\alpha =$ 2.35 and 0.
If the upper limiting mass in the star formation is considerably large, say, $M_{\rm up} > 430 M_\odot$,
then the relative event rate is nearly independent from $f_{\rm cag}$.
This constant value becomes $\sim1$\% for the Salpeter IMF case.
Also, the relative rate only slightly depends on $M_{\rm up}$ for $f_{\rm cag} = 1.2$ case.
On the other hand, the relative event rate shows a $M_{\rm up}$ dependency 
for models with $f_{\rm cag} < 1.2$.
In particular, no PISN progenitor is formed under the condition of
small $M_{\rm up}$ and small $f_{\rm cag}$.
Considering the big uncertainty in the current estimate of $M_{\rm up}$,
we conclude that
the combination of the small $M_{\rm up}$ and the large carbon-to-oxygen ratio 
has a potential importance to explain the nondetection of PISNe.

\subsection{Observational consequences}

The luminosity of a PISN in the early phase will be powered by
diffusion of the thermal energy that is deposited by the shock heating and
the radioactive decay of $^{56}$Ni $\rightarrow$ $^{56}$Co.
The duration is determined by the diffusion time,
which scales as $t_{\rm diff} \sim 2\times10^6 ~ {\rm sec} \
(M_{\rm ej}/M_\odot)^{3/4}
(E_{\rm exp}/10^{51} ~ {\rm erg})^{-1/4}
(\kappa / 0.4 ~ {\rm cm}^2 ~ {\rm g}^{-1})^{1/2}$,
where $M_{\rm ej}$, $E_{\rm exp}$, and $\kappa$ are
the ejecta mass, the explosion energy, and the opacity, respectively \citep{arnett80}.
As the opacity in a PISN is dominated by the electron scattering
and thus $\kappa \sim 0.4 ~ {\rm cm}^2 ~ {\rm g}^{-1}$ \citep{kasen+11},
the large ejecta mass of a PISN of $\sim 100 ~ M_\odot$ results in
a long diffusion time of $\gtrsim 200$ day, which characterizes the PISN light curve.
The peak luminosity during the diffusion phase will correlate with the explosion energy \citep[c.f.][]{kasen+11}.
The late luminosity, which is explained by the $^{56}$Co decay, is estimated as
$L_{\rm dec}(t) \sim 1.63 \times 10^{43} (M_{\rm ^{56}Co}/M_\odot) \exp( -t/9.60\times10^6 ~ {\rm sec})
~{\rm erg}~{\rm s}^{-1}$ \citep{arnett79},
where $M_{\rm ^{56}Co}$ is the mass of the $^{56}$Co,
which is originally ejected as $^{56}$Ni.

Therefore, a PISN can have a less luminous brightness,
if the explosion energy and the ejected $^{56}$Ni mass are small.
Indeed, the model R175 in \citet{kasen+11},
which is a 175 $M_\odot$ PISN model producing
$21.3 \times 10^{51}$ erg of the explosion energy and 0.70 $M_\odot$ of $^{56}$Ni,
has a peak absolute R-band magnitude of $\sim -17$ and a fainter decay tail
that are rather comparable to a normal Type IIP SNe.
Our minimum mass PISN models for various $f_{\rm cag}$ will have
similar observational properties to this R175 model,
because they also have $\sim 20 \times 10^{51}$ erg of the explosion energies
and $\sim 1$ $M_\odot$ $^{56}$Ni yields.
Therefore, the most important observational consequence of the high $X$(C)/$X$(O)
is the more fainter PISN from the same mass progenitors.
We expect that the low detectability of PISNe may be explained by
the fainter luminosity of the VMS progenitors with higher $X$(C)/$X$(O).

Apart from the discussion on the low detectability,
we discuss some possibilities to constrain the $X$(C)/$X$(O) in the real VMSs.
For example, the core carbon-to-oxygen ratio can be constrained
by determining the relation between the ejecta mass and the explosion energy.
This relation can be obtained from the width of the light curve,
if the explosion energy is determined by 
another kind of observation, such as the line broadening.
Similarly, the relation between the ejecta mass and the $^{56}$Ni yield also
significantly depends on the $X$(C)/$X$(O).
By observing the late decay tail of the light curve,
the $^{56}$Ni ejecta mass can be obtained.
This will also be a powerful tool to distinguish models
with different core carbon-to-oxygen ratios.

\begin{figure}[t]
	\centering
	\includegraphics[width=0.5\textwidth]{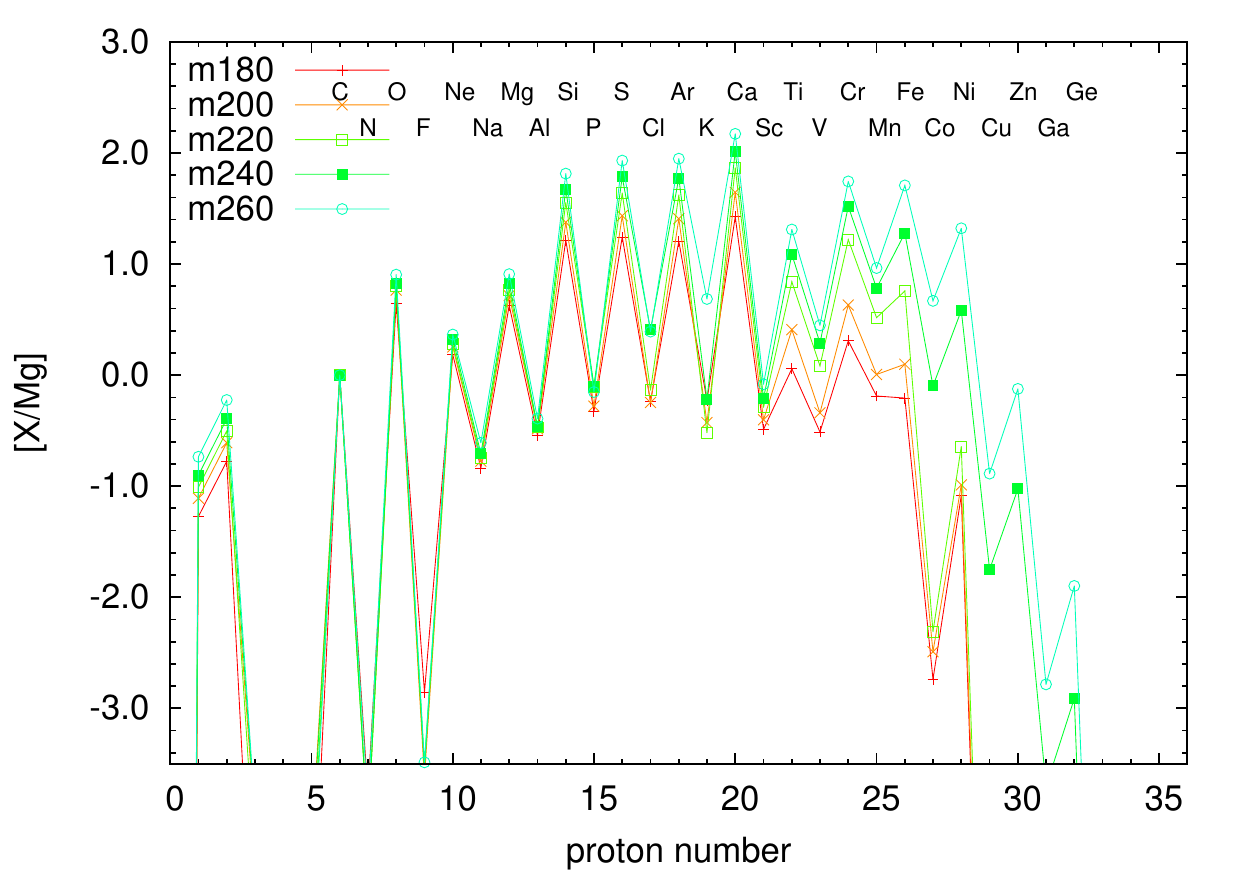}
	\includegraphics[width=0.5\textwidth]{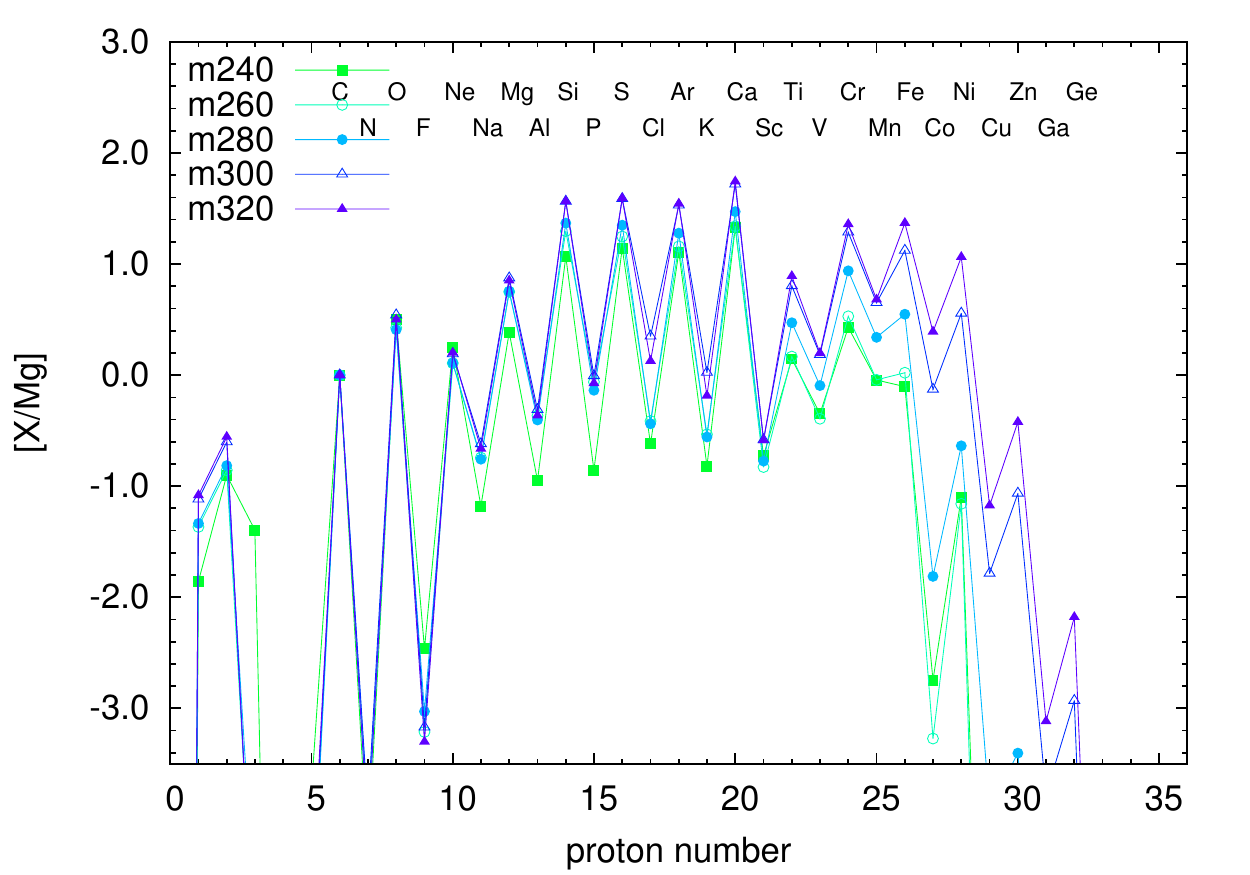}
	\caption{\footnotesize{The abundance ratios of PISN explosive yields
	for models with $f_{\rm cag}=1.2$ (top) and 0.6 (bottom).}}
	\label{fig-abundance}
\end{figure}

On the other hand, no significant distinction is found
from PISN yields with different $X$(C)/$X$(O),
in spite of the big difference in the initial mass ranges.
The abundance ratios of PISN explosive yields are shown 
in Fig. \ref{fig-abundance} for models with $f_{\rm cag}=0.6$ and 1.2,
in which the solar values in \citet{asplund+09} are used.
The low [Na/Mg] and the high [Ca/Mg],
which characterize the PISN yields from ordinary CCSN yields \citep{takahashi+18a},
are resulted from both of the sequences.
The explosive yields are incapable to discriminate
stellar models with different core carbon-to-oxygen ratio.

\section{Conclusion}

Thanks to the development of automated wide-field surveys,
currently more than 1,000 of SNe are discovered every year.
The large number might be enough for the detection of PISNe,
because the relative event rate of $\sim$1\% of CCSN events is estimated
from the conventional stellar evolution simulations for the Salpeter initial mass function.
However, none of the observed SNe are known to show
characteristic signatures of PISNe such as the intrinsically red color
and the broad light curve.

The above estimate of $\sim1$\% of the relative event rate of PISNe to CCSNe is
based on a conventional estimate for the initial mass range of PISNe of $\sim$140--260 $M_\odot$.
Because more massive stars are less frequently formed in the present universe,
the event rate of PISNe has been possibly overestimated 
if the upper and lower ends of the PISN mass range 
has been underestimated than their actual values.
So far, most estimates of the PISN initial mass range have assumed
the well-defined mass range of the CO core for PISNe of $\sim$65--120 $M_\odot$.
For example, a strong wind mass loss has been known to affect the PISN event rate
by shifting the initial mass range for PISNe upward
for VMSs with finite metallicities but without changing the CO core mass range.

In this work, we have investigated the VMS evolution with various core carbon-to-oxygen ratios.
By applying a modulation factor of $f_{\rm cag} \in [0.1, 1.2]$
to the reaction rate of $^{12}$C($\alpha$,$\gamma$)$^{16}$O of \citet{Caughlan&Fowler88},
VMS models developing CO cores with 
$X(\rm{C})/X(\rm{O}) \sim$ 0.15--3.1 have been calculated.
The characteristic excited states of the compound nuclei $^{16}$O makes it challenging 
to accurately determine the $^{12}$C($\alpha$,$\gamma$)$^{16}$O reaction rate \citep{deBoer+17}.
Although the small reaction rate of \citet{Caughlan&Fowler88}
is below the uncertainty of the most recent estimates \citep{Xu+13, deBoer+17},
modulation factors of $f_{\rm cag} \gtrsim 0.8$ is compatible with the estimate of \citet{Buchmann96}.
Moreover, the high core carbon fraction may result from 
astrophysical mechanisms such as the additional mixing,
because the mixing at the convective boundary region
during the last part of the core He burning phase
can significantly affect the carbon abundance in the convective core \citep{Imbriani+01}.
Thus it is still interesting to investigate what results from
the high carbon fraction in CO cores formed in VMSs.

We have found that VMSs with high core carbon-to-oxygen ratios
follow a qualitatively different evolutionary path from conventional models.
Less massive models with small $f_{\rm cag}$
avoid the pair-creation instability, since their effective core masses 
are reduced during the carbon-burning phase by developing shell convection.
For example, this takes place for $M_{\rm ini} <$ 105, 135, and 155 $M_\odot$ models
for $f_{\rm cag} =$ 1.0, 0.8, and 0.6 cases, respectively.
Besides, massive exploding models with smaller $f_{\rm cag}$ are found to have higher explodabilities,
i.e., stars with a higher core carbon-to-oxygen ratio explode with smaller explosion energies.
For example, the explosion energies of 260 $M_\odot$ models are
73.7, 65.1, and 44.3 $\times 10^{51}$ erg
for $f_{\rm cag} =$ 1.0, 0.8, and 0.6 cases, respectively.
Consequently, the initial mass range for PISNe increases from 
$M_{\rm ini} \in [ 175, 270 ]$ $M_\odot$ for the conservative $f_{\rm cag} = 1.2$ case to
$M_{\rm ini} \in [ 210, 290 ]$ $M_\odot$,
$M_{\rm ini} \in [ 210, 310 ]$ $M_\odot$, and
$M_{\rm ini} \in [ 230, 330 ]$ $M_\odot$ for 
$f_{\rm cag} = $1.0, 0.8, and 0.6 cases, respectively.
It has been also found that, as well as the explosion energy, 
the $^{56}$Ni yield significantly decreases with decreasing $f_{\rm cag}$.

We have estimated the corresponding relative event rate of PISNe to that of CCSNe
by integrating a simplified IMF that is characterized by the slope, $\alpha$,
and the upper limiting mass for the star formation, $M_{\rm up}$.
With sufficiently large $M_{\rm up} > 430$ M$_\odot$, 
the relative rate becomes nearly independent from $f_{\rm cag}$, and
a roughly constant value of $\sim$1\% is obtained
for the Salpeter value of $\alpha = 2.35$.
The event rate can be significantly reduced by decreasing $M_{\rm up}$, and
the reduction is more vigorous for models with smaller $f_{\rm cag}$, 
or with higher core carbon-to-oxygen ratios.
This result advances the first theory to decrease the PISN event rate
not by modifying the initial mass--CO core mass relation
but by directly changing the CO core mass range.

Finally, observational consequences of PISNe
with different core carbon-to-oxygen ratios are discussed.
Based on the small explosion energies and the small $^{56}$Ni yields,
the minimum mass PISNe for different $f_{\rm cag}$ cases 
are estimated to have similar luminosities to a normal Type IIP SN.
Therefore, those relatively fainter PISNe may be missed from extensive observations,
explaining the low detectability of the PISNe.\\

The author thanks Prof. Nobert Langer and Dr. Takashi Yoshida for fruitful discussions.
The author is grateful to the anonymous referee for his/her careful reading of the manuscript and helpful comments.
K. T. was supported by Japan Society for the Promotion of Science (JSPS) Overseas Research Fellowships.

%%%%%%%%%%%%%%%%%%%% REFERENCES %%%%%%%%%%%%%%%%%%
\bibliography{biblio}

\end{document}